\documentclass[a4paper]{jpsj3}
\usepackage{bm}% bold math

\newcommand{\beq}{\begin{equation}}
\newcommand{\eeq}{\end{equation}}
\newcommand{\bea}{\begin{eqnarray}}
\newcommand{\eea}{\end{eqnarray}}
\newcommand{\bec}{\begin{center}}
\newcommand{\enc}{\end{center}}
\newcommand{\bfr}{\begin{flushright}}
\newcommand{\efr}{\end{flushright}}
\newcommand{\la}{\langle}
\newcommand{\ra}{\rangle}

\newcommand{\alp}{\alpha}

\newcommand{\om}{\omega}
\newcommand{\kap}{\kappa}
\newcommand{\g}{\gamma}
\newcommand{\s}{\sigma}

\newcommand{\lam}{\lambda}

\newcommand{\Res}{\mathop{\rm Res}\nolimits}

\newcommand{\dint}{\int\hspace{-2.5mm}\int}
\newcommand{\tb}{\widetilde{b}}%\rm 
\newcommand{\tc}{\widetilde{c}}%\rm 
\newcommand{\td}{\widetilde{d}}%\rm 
\newcommand{\tom}{\widetilde{\omega}}

\newcommand{\cH}{{\cal H}}

\newcommand{\cL}{{\cal L}} 
\newcommand{\cN}{{\cal N}} 
\newcommand{\cP}{{\cal P}}

\newcommand{\gammap}{\gamma_p}

\title{Properties of a Single Photon Generated by a Solid-State Emitter: \\
Effects of Pure Dephasing}

\author{
\name{Eiki \surname{Iyoda}$^{1}$\footnote{Present address: Department of Physics, Tohoku University, 
Sendai 980-8578, Japan}},
\name{Takeo \surname{Kato}$^{2}$},
\name{Takao \surname{Aoki}$^{3}$},
\name{Keiichi \surname{Edamatsu}$^{4}$},
and \name{Kazuki \surname{Koshino}$^{1}$}\thanks{E-mail address: kazuki.koshino@osamember.org}
}

\inst{
$^1$College of Liberal Arts and Sciences, 
Tokyo Medical and Dental University, Ichikawa, Chiba 272-0827, Japan \\
$^2$Institute for Solid State Physics, the University of Tokyo, Kashiwa, Chiba 277-8581, Japan\\
$^3$Department of Applied Physics, Waseda University, Tokyo 169-8555, Japan \\
$^4$Research Institute of Electrical Communication, Tohoku University, Sendai 980-8577, Japan\\
}

\abst{
We investigate the properties of a single photon 
generated by a solid-state emitter
subject to strong pure dephasing.
We employ a model in which all the elements of the system,
including the propagating fields, are treated quantum mechanically.
We analytically derive the density matrix of the emitted photon,
which contains full information about the photon,
such as its pulse profile, power spectrum, and purity. 
We visualize these analytical results using realistic parameters
and reveal the conditions for maximizing the purity of generated photons.
}

\kword{single-photon source, pure dephasing, purity}

\begin{document}
\maketitle
%%%%%%%%%%%%%%%%%%%%%%%%%%%%%%%%%%%%%%%%%%%%%%%%%%%%%%%%%%%%%%%%%%%%%%%%%%%%%%%
\section{Introduction}
%%%%%%%%%%%%%%%%%%%%%%%%%%%%%%%%%%%%%%%%%%%%%%%%%%%%%%%%%%%%%%%%%%%%%%%%%%%%%%%
Cavity quantum electrodynamics (QED)
is currently one of the hottest research topics in atomic physics. In particular, 
quantum-mechanical interactions between atoms and cavity photons
have been intensively studied~\cite{Miller05}.
Solid-state cavity QED systems
composed of semiconductor quantum dots 
and cavities have recently been attracting much attention
since they are suitable for creating 
compact optical devices~\cite{Khitrova06}.
Strong coupling between a single dot and a cavity has been confirmed
through a large vacuum Rabi splitting~\cite{Reithmaler04,Yoshie04,Peter05}.
This strong coupling has been applied to fabricate a single-dot laser~\cite{Strauf06}
and to generate nonclassical light
including single photons~\cite{Srinivasan07a,Srinivasan07b,Michler00,Santori01,Moreau01}.
Excellent performances have been reported
in generating indistinguishable photons~\cite{Santori02,Varoutsis05} 
and entangled photon pairs~\cite{Fattal04}, 
both of which are useful for quantum information processing~\cite{Kiraz04}.

In contrast to real atoms,
semiconductor quantum dots are strongly influenced by 
environmental noise sources such as phonons and background carriers.
The fluorescence spectra of solid-state and atomic cavity QED systems are qualitatively different. When a solid-state system is excited by pump light of the dot frequency,
a spectral peak appears at the cavity frequency 
in spite of the large detuning between them~\cite{Reithmaler04,Yoshie04,Peter05}.
This phenomenon has not been reported for atomic cavity QED systems.
Further experimental studies have characterized this peak more 
thoroughly~\cite{Hennessy07,Muller07,Press07,Suffczynski09,Ota09,Ates09,Ulhaq10}
and have revealed that the fluorescence at the cavity frequency 
is due to radiative decay of the dot.
Subsequent theoretical studies accounted for the pure dephasing of the dot 
through the stochastic Schr\"odinger equation~\cite{Cui06,Naesby08} 
or the Master equation~\cite{Yamaguchi08,Auffeves09,Auffeves10} 
and successfully explained the peak at the cavity frequency.
The influence of pure dephasing on the radiative decay of the dot
can be understood in terms of the quantum Zeno and anti-Zeno 
effects~\cite{Koshino05,Yamaguchi08}.

Therefore, when designing a single-photon source
using solid-state cavity QED systems,
it is crucial to quantitatively consider the pure dephasing of the dot. 
The performance of such photon sources
should be evaluated from two aspects.
One is the collection efficiency, namely, 
the probability that the emitted photon
is transferred to the intended spatial mode (i.e., the radiation pattern of the cavity).
This has been discussed in several studies 
in terms of the ratio of radiative decay rates~\cite{Cui06,Auffeves09,Auffeves10}. 
The other is the indistinguishability of generated photons,
which can be measured by two-photon interference experiments
and is evaluated by the purity.
Single photons with high purity are required for quantum information processing,
particularly for constructing scalable quantum circuits.

In this study, we investigate the properties of a single photon
emitted by a solid-state cavity QED system
and quantitatively observe the effects of pure dephasing.
In order to obtain full information including the indistinguishability of generated single photons,
we treat the five elements of the overall system
(the dot, the cavity, radiation leaking from the cavity, 
non cavity radiation modes,
and the environment causing the pure dephasing of the dot)
as active quantum-mechanical degrees of freedom,
and analytically derive the density matrix of the emitted photon
in the real-space representation.
This density matrix contains full information about the emitted photon,
including its pulse profile, frequency spectrum, and purity.
These quantities are observed as functions of the pure dephasing rate of the dot.
We reveal the optimum condition for maximizing the purity of the emitted photon.

%%%%%%%%%%%%%%%%%%%%%%%%%%%%%%%%%%%%%%%%%%%%%%%%%%%%%%%%%%%%%%%%%%%%%%%%%%%%%%%
\section{Model}
\label{sec:model}
%%%%%%%%%%%%%%%%%%%%%%%%%%%%%%%%%%%%%%%%%%%%%%%%%%%%%%%%%%%%%%%%%%%%%%%%%%%%%%%
%-------------------------------------------------------------------------------
\begin{figure}[t]
\begin{center}
\includegraphics[width=0.4 \columnwidth]{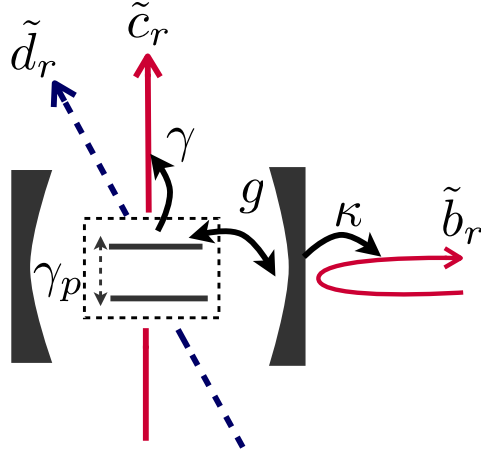}
\caption{(Color online) Schematic illustration of the solid-state cavity QED system considered.
It consists of a quantum dot, a cavity,
photon leakage from the cavity ($b$ field),
non-cavity radiation modes ($c$ field),
and a reservoir field, which causes the pure dephasing of the dot ($d$ field).}
\label{fig:1}
\end{center}
\end{figure}
%-------------------------------------------------------------------------------

We investigate the radiative decay of an excited quantum dot placed inside a cavity,
as illustrated in Fig.~\ref{fig:1}.
This solid-state cavity QED system consists of the following five components:
(i)~a quantum dot,
(ii)~a cavity,
(iii)~a photon field leaking from the cavity (referred to as $b$ field hereafter),
(iv)~non-cavity radiation modes ($c$ field), and 
(v)~a reservoir field, which causes the pure dephasing of the dot ($d$ field)~\cite{recentPRA}.
The annihilation operators corresponding to these components 
are respectively denoted as $\sigma$, $a$, $b_k$, $c_k$, and $d_k$,
where $k$ is a one-dimensional wave number.
Note that $\sigma$ is a Pauli operator, whereas the other operators are bosonic.
Setting $\hbar = c = 1$, the Hamiltonian of the overall system is given by
\begin{align}
\cH &= \cH_0+\cH_1+\cH_2+\cH_3, 
\label{eq:Ham} \\
{\cal H}_0 &= \omega_d\sigma^\dag\sigma + \omega_c a^\dag a + g(\sigma^\dag a+a^\dag\s), 
\label{eq:Ham0} \\
{\cal H}_1 &= \int dk 
\left[
kb^\dag_k b_k + \sqrt{\kappa/(2\pi)}(a^\dag b_k+b_k^{\dagger}a) 
\right],
\label{eq:Ham1} \\
{\cal H}_2 &= \int dk 
\left[kc^\dag_k c_k 
+ \sqrt{\gamma/(2\pi)}(\sigma^\dag c_k+c_k^{\dag}\sigma)
\right],
\label{eq:Ham2} \\
{\cal H}_3 &= \int dk \left[
k d^\dag_k d_k 
+ \sqrt{\gamma_p/\pi}\sigma^{\dagger}\sigma
(d_k^{\dagger}+d_k)
\right].
\label{eq:Ham3} 
\end{align}
The parameters are defined as follows (see Fig.~\ref{fig:1}).
$\om_d$ and $\om_c$ respectively denote
the resonance frequencies of the dot and cavity,
$g$ represents the coupling between them,
$\kap$ is the escape rate of cavity photons,
$\gamma$ is the radiative decay rate of the dot into non-cavity modes,
and $\gamma_p$ is the pure dephasing rate of the dot.
$\cH_0$ describes the Rabi oscillation between the dot and the cavity
(Jaynes--Cummings Hamiltonian),
$\cH_1$ describes the leakage of a cavity photon to its radiation pattern,
$\cH_2$ describes the radiative decay of the dot in unintended directions,
and $\cH_3$ describes the pure dephasing of the dot.
We can confirm that 
${\cal N}\equiv\sigma^{\dagger}\sigma + a^{\dagger}a 
+ \int dr \tb^{\dagger}_r \tb_r + \int dr \tc^{\dagger}_r \tc_r$
commutes with the Hamiltonian.
Therefore, the number of excitations is conserved
in the dot, cavity, and $b$ and $c$ fields.

We assume that the dot is initially ($t=0$)
in the excited state
while the other fields are in their vacuum states.
Then, denoting the overall vacuum state by $|0\ra$,
the initial state vector is given by
\beq
|\psi_i\ra = \sigma^\dag |0\ra.
\label{eq:init}
\eeq
The Hamiltonian of eq.~(\ref{eq:Ham}) and 
the initial state vector of eq.~(\ref{eq:init})
form the basis of our analysis.

For later convenience, we introduce the real-space representation 
of the $b$ field (cavity leakage). It is defined by
\beq
\tb_r = (2\pi)^{-1/2} \int dk e^{ikr}b_k. 
\label{eq:tbr}
\eeq
The $r<0$ ($r>0$) region represents the incoming (outgoing) field.
$\tc_r$ and $\td_r$ can be formally defined in a similar manner.
Our main concern lies in the properties of a single photon emitted in the $b$ field.

To model the pure dephasing of the dot,
the $d$ field interacts with the dot
so as to conserve the dot excitation.
Using eqs.~(\ref{eq:Ham0}) and (\ref{eq:Ham3}), 
the dot Hamiltonian can be rewritten as $[\omega_d + f(t)]\sigma^{\dagger} \sigma$, 
where $f(t) = \sqrt{2\gammap}[\td_0(t) + \td_0^{\dagger}(t)]$ is 
the fluctuation of the dot resonance frequency induced by the $d$ field.
Using eqs.~(\ref{eq:init}) and (\ref{InputOutput3}), 
we can confirm that $\la f(t)f(t') \ra_i = 2\gammap \delta(t-t')$,
where $\la \cdots\ra_i=\la\psi_i|\cdots|\psi_i\ra$.
Therefore, the present model assumes a white noise spectrum 
for the fluctuation of the dot resonance.

%%%%%%%%%%%%%%%%%%%%%%%%%%%%%%%%%%%%%%%%%%%%%%%%%%%%%%%%%%%%%%%%%%%%%%%%%%%%%%%
\section{Analysis}
%%%%%%%%%%%%%%%%%%%%%%%%%%%%%%%%%%%%%%%%%%%%%%%%%%%%%%%%%%%%%%%%%%%%%%%%%%%%%%%
In this section, we present analytical results
that are rigorously derivable from the model described in \S\ref{sec:model}.
We solve the time evolution of the overall system within the input-output formalism~\cite{Walls95,Gardiner04}
and derive several formulae to characterize the emitted single photon
(density matrix, pulse shape, spectrum, and purity).
These analytical results are visualized in the next section
for specific parameters.

%%%%%%%%%%%%%%%%%%%%%%%%%%%%%%%%%%%%%%%%%%%%%%%%%%%%%%%%%%%%%%%%%%%%%%%%%%%%%%%
\subsection{Heisenberg equations}
%%%%%%%%%%%%%%%%%%%%%%%%%%%%%%%%%%%%%%%%%%%%%%%%%%%%%%%%%%%%%%%%%%%%%%%%%%%%%%%
Here, we present the Heisenberg equations for 
the system ($\s$, $a$) and field ($b_k$, $c_k$, $d_k$) operators.
Deriving the raw Heisenberg equations for the field operators 
from eq.~(\ref{eq:Ham}) and transforming them into real-space representations,
we obtain the following relations
that connect the incoming ($r<0$) and outgoing ($r>0$) fields:
\begin{align}
\tb_r(t) &= \tb_{r-t}(0)-i\sqrt{\kappa}\theta(r)\theta(t-r)a(t-r),
\label{InputOutput1} \\
\tc_r(t) &= \tc_{r-t}(0)-i\sqrt{\gamma}\theta(r)\theta(t-r)\sigma(t-r),
\label{InputOutput2} \\
\td_r(t) &= \td_{r-t}(0)-i\sqrt{2\gammap}\theta(r)\theta(t-r)\sigma^\dag(t-r)\sigma(t-r),
\label{InputOutput3} 
\end{align}
where $\theta(x)$ is the Heaviside step function.
From the raw Heisenberg equations for the system operators
and the above input--output relations,
the Heisenberg equations for system operators are given by
\begin{align}
\frac{d}{dt} \sigma 
&= -i\tom_d\sigma -ig(1-2\sigma^\dag\sigma)a
-i(1-2\sigma^\dag\sigma)N_c(t)
-i\left[N_d^{\dagger}(t)\s + \s^{\dag}N_d(t)\right],
\label{HE1} \\
\frac{d}{dt}a 
&=-i\tom_c a -ig\sigma -iN_b(t),
\label{HE2}
\end{align}
where $\tom_d=\om_d-i(\gamma/2+\gammap)$ and $\tom_c=\om_c-i\kap/2$ 
respectively are the complex frequencies of the dot and cavity,
and the noise operators are defined by
$N_b(t)=\sqrt{\kap}\tb_{-t}(0)$,
$N_c(t)=\sqrt{\gamma}\tc_{-t}(0)$, and 
$N_d(t)=\sqrt{2\gammap}\td_{-t}(0)$. 
Note that the noise operators are the initial-time operators
and, consequently, $N_j(t)|0\ra=0$ ($j=b, c, d$).

%%%%%%%%%%%%%%%%%%%%%%%%%%%%%%%%%%%%%%%%%%%%%%%%%%%%%%%%%%%%%%%%%%%%%%%%%%%%%%%
\subsection{State vector}
\label{sec:sv}
%%%%%%%%%%%%%%%%%%%%%%%%%%%%%%%%%%%%%%%%%%%%%%%%%%%%%%%%%%%%%%%%%%%%%%%%%%%%%%%
The state vector of the overall system
at an arbitrary time $t$ is determined by $|\psi(t)\ra = e^{-i\cH t}|\psi(0)\ra$.
Since the initial dot excitation is conserved 
in the dot, cavity, and $b$ and $c$ fields, 
the state vector can be written as
\begin{align}
\left|\psi(t)\right> &= 
\left[ \alpha_0(t)\sigma^\dag + \beta_0(t)a^\dag
+ \int dr\gamma_0(r,t)\tb_r^\dag + \int dr \delta_0(r,t) \tc_r^{\dagger}\right]|0\ra
 \nonumber \\
& + \sum_{m=1}^{\infty} \int d^m{\bm x} 
\left[ \alpha_m({\bm x},t) \sigma^\dag 
+ \beta_m({\bm x},t) a^\dag + \int dr \gamma_m(r,{\bm x},t) \tb^\dag_r 
+ \int dr \delta_m(r,{\bm x},t) \tc^\dag_r \right] 
\td_{x_1}^{\dagger} \cdots \td_{x_m}^{\dagger}|0\ra, 
\label{eq:vec}
\end{align}
where $m$ denotes the number of excitations in the $d$ field and 
$\int d^m{\bm x}$ denotes a multi-dimensional integral 
with respect to ${\bm x} = (x_1, x_2, \cdots, x_m)$.
We can set $x_1 \le \cdots \le x_m$ without loss of generality.
As we show later, these coefficients are nonzero only when
$0\le r \le x_1 \le \cdots \le x_m \le t$.

\begin{figure}[t]
\begin{center}
\includegraphics[width=0.5\columnwidth]{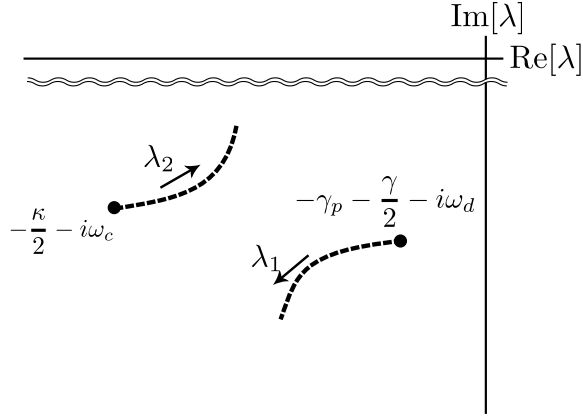}
\caption{$\lambda_1$ and $\lambda_2$ in the complex plane.
$\lambda_1=-i\tom_d$ and $\lambda_2=-i\tom_c$ when $g=0$.
Dotted lines show their traces as $g$ increases.}
\label{fig:lam12}
\end{center}
\end{figure}

First, we discuss $\alp_0$ and $\beta_0$.
From eq.~(\ref{eq:vec}), we can confirm that
$\alp_0(t)=\la\s(t)\s^{\dag}(0)\ra$ and 
$\beta_0(t)=\la c(t)\s^{\dag}(0)\ra$,
where $\la\cdots\ra=\la 0|\cdots|0 \ra$.
From eqs.~(\ref{HE1}) and (\ref{HE2}),
their equations of motion are given by
\begin{align}
\frac{d}{dt} \begin{pmatrix} \alpha_0(t) \\ \beta_0(t) \end{pmatrix} &= 
\begin{pmatrix} 
-i\tom_d & -ig \\
-ig & -i\tom_c
\end{pmatrix}
\begin{pmatrix}
\alpha_0(t) \\
\beta_0(t) 
\end{pmatrix},
\label{eq:a0b0}
\end{align}
with the initial conditions $\alpha_0(0)=1$ and $\beta_0(0)=0$. 
The solutions are given by
\begin{align}
\alpha_0(t) &= A_1e^{\lambda_1t} + A_2e^{\lambda_2t}, \\
\beta_0(t) &= B_1 e^{\lambda_1t} + B_2e^{\lambda_2t},
\end{align}
where $\lambda_1$ and $\lambda_2$ are the two eigenvalues
of the $2\times 2$ matrix in eq.~(\ref{eq:a0b0})
(see Fig.~\ref{fig:lam12}),
$A_1=(\lambda_1+i\tom_c)/(\lambda_1-\lambda_2)$,
$A_2=(\lambda_2+i\tom_c)/(\lambda_2-\lambda_1)$,
and $B_1=-B_2=-ig/(\lambda_1-\lambda_2)$.
The real parts of $\lambda_1$ and $\lambda_2$ are always negative
and, consequently, $\alp_0$ and $\beta_0$ vanish as $t\to\infty$.
Equation~(\ref{eq:vec}) also implies that
$\gamma_0(r,t)=\la \tb_r(t)\s^{\dag}(0)\ra$ and 
$\delta_0(r,t)=\la \tc_r(t)\s^{\dag}(0)\ra$.
From eqs.~(\ref{InputOutput1}) and (\ref{InputOutput2}), we have
\begin{align}
\gamma_0(r,t) & = -i\sqrt{\kappa}\beta_0(t-r), \\
\delta_0(r,t) & = -i\sqrt{\gamma}\alpha_0(t-r).
\end{align}
Rigorously, $\theta(r)\theta(t-r)$ should appear 
on the right-hand sides of these equations.
However, below,
we implicitly assume $0\le r \le x_1 \le \cdots \le x_m \le t$
and omit the Heaviside functions.

%-------------------------------------------------------------------------------
%-------------------------------------------------------------------------------

Next, we proceed to investigate higher-order quantities including reservoir ($d$ field) excitations.
We consider $\alp_1$ and $\beta_1$ as examples.
Applying the same reasoning as that for $\gamma_0$ and $\delta_0$, we have 
$\alp_1(x,t)=-i\sqrt{2\g_p}\la\s(t)\s^{\dag}(t-x)\s(t-x)\s^{\dag}(0)\ra$ and 
$\beta_1(x,t)=-i\sqrt{2\g_p}\la a(t)\s^{\dag}(t-x)\s(t-x)\s^{\dag}(0)\ra$.
Thus, we must evaluate the three-time correlation functions
$\la\s(t_2)\s^{\dag}(t_1)\s(t_1)\s^{\dag}(0)\ra$ and 
$\la a(t_2)\s^{\dag}(t_1)\s(t_1)\s^{\dag}(0)\ra$
with $t_2>t_1>0$.
Their equations of motion with respect to $t_2$
have the same form as eq.~(\ref{eq:a0b0})
and their initial values ($t_2\to t_1$) are $\alp_0(t_1)$ and $0$, respectively.
This implies that 
the two-time correlation functions can be factorized as
$\la\s(t_2)\s^{\dag}(t_1)\s(t_1)\s^{\dag}(0)\ra=\alp_0(t_2-t_1)\alp_0(t_1)$ and 
$\la a(t_2)\s^{\dag}(t_1)\s(t_1)\s^{\dag}(0)\ra=\beta_0(t_2-t_1)\alp_0(t_1)$.
Thus, we have
\begin{align}
\alpha_1(x,t) &= -i\sqrt{2\g_p}\alp_0(t-x)\alp_0(x), \\
\beta_1(x,t)  &= -i\sqrt{2\g_p}\alp_0(t-x)\beta_0(x).
\end{align}
Repeating the same arguments, all coefficients can be written 
as products of $\alp_0$ and $\beta_0$, as follows:
\begin{align}
\alpha_m({\bm x},t) &= \left(-i\sqrt{2\gammap}\right)^m 
K_m({\bm x},t)\alp_0(x_1),
\label{eq:alpm}
\\
\beta_m({\bm x},t) &= \left(-i\sqrt{2\gammap}\right)^m 
K_m({\bm x},t)\beta_0(x_1),
\\
\gamma_m(r,{\bm x},t)  &= \left(-i\sqrt{\kappa}\right)\left(-i\sqrt{2\gammap}\right)^m 
K_m({\bm x},t)\beta_0(x_1-r),\label{eq:FactorizeGamma}
\\
\delta_m(r,{\bm x},t) &= \left(-i\sqrt{\gamma}\right)\left(-i\sqrt{2\gammap}\right)^m
K_m({\bm x},t)\alp_0(x_1-r),
\end{align}
where
\begin{align}
K_m({\bm x},t) &= \alp_0(t-x_m)\alp_0(x_m-x_{m-1})\cdots\alp_0(x_2-x_1).
\end{align}

%%%%%%%%%%%%%%%%%%%%%%%%%%%%%%%%%%%%%%%%%%%%%%%%%%%%%%%%%%%%%%%%%%%%%%%%%%%%%%%
\subsection{Decay of dot excitation}
\label{ssec:decay}
\begin{figure}[t]
\begin{center}
\includegraphics[width=0.5\columnwidth]{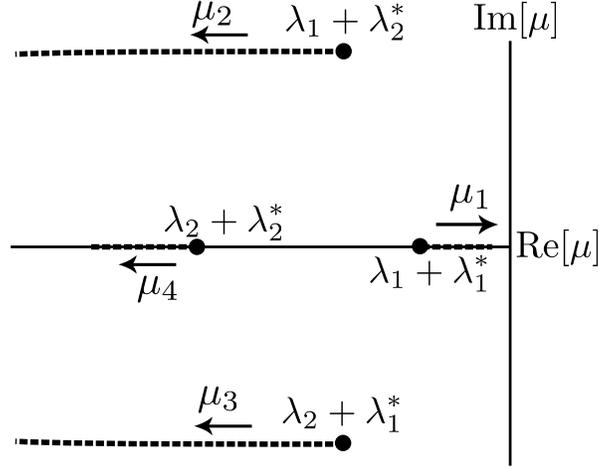}
\caption{
$\mu_j$ ($j=1,\cdots,4$) in the complex plane.
When $\gamma_p$ is absent, $\mu_1=\lam_1+\lam_1^*$,
$\mu_2=\lam_1+\lam_2^*$, $\mu_3=\lam_2+\lam_1^*$, and
$\mu_4=\lam_2+\lam_2^*$.
Dotted lines indicate their traces as $\gamma_p$ increases.
Real parts of $\mu_j$ are always negative for any $\gamma_p$.}
\label{fig:muj}
\end{center}
\end{figure}

%%%%%%%%%%%%%%%%%%%%%%%%%%%%%%%%%%%%%%%%%%%%%%%%%%%%%%%%%%%%%%%%%%%%%%%%%%%%%%%
The state vector of eq.~(\ref{eq:vec})
fully describes the dynamics of the overall system,
including both its transient and asymptotic behaviors.
In this section, as an example of a transient phenomenon, we analyze the decay of dot excitation.
The survival probability of dot excitation is defined as
$P(t)=\la\psi(t)|\s^{\dag}\s|\psi(t)\ra$.
From eqs.~(\ref{eq:vec}) and (\ref{eq:alpm}), we have
\begin{align}
P(t) &= |\alpha_0(t)|^2 + 2\gammap\int dx |\alpha_0(t-x)|^2|\alpha_0(x)|^2 +\cdots.
\end{align}
We here introduce the Laplace transform of $|\alp_0|^2$, which is
defined by $\cL_{|\alp_0|^2}(z)=\int_0^{\infty}dt e^{-zt} |\alpha_0(t)|^2$.
It is given by
\begin{align}
\cL_{|\alp_0|^2}(z) 
= \sum_{m,n=1,2} \frac{A_m A_n^{\ast}}{z-\lambda_m -\lambda_n^{\ast}},
\end{align}
where $\lam_{1,2}$ and $A_{1,2}$ are defined in \S\ref{sec:sv}.
The Laplace transform of $P(t)$ is then given by
\begin{align}
\cL_{P}(z) 
&= \frac{\cL_{|\alp_0|^2}(z)}{1-2\gammap \cL_{|\alp_0|^2}(z)}.
\end{align}
$P(t)$ is obtained by analyzing 
the poles of this function in the $z$-plane.
We denote the four roots of the equation $1-2\gammap \cL_{|\alp_0|^2}(z)=0$
by $\mu_j$ ($j=1,\cdots,4$) (see Fig.~\ref{fig:muj}).
$P(t)$ is then given by
\begin{align}
P(t) &= \sum_{j=1}^{4} E_j e^{\mu_j t}, \label{AnalyticSP} \\
E_j &= \frac{\prod_{m',n'=1,2}(\mu_j-\lambda_{m'}-\lambda^{\ast}_{n'})}
{\prod_{i(\ne j)}(\mu_j-\mu_i)}
\sum_{m,n=1,2} \frac{A_m A_n^{\ast}}{\mu_j-\lambda_m-\lambda_n^{\ast}}.
\end{align}
Note that the real parts of $\mu_j$ are always negative and that
the survival probability $P(t)$ vanishes
in the $t\to\infty$ limit, as expected.

%-------------------------------------------------------------------------------
%-------------------------------------------------------------------------------

%%%%%%%%%%%%%%%%%%%%%%%%%%%%%%%%%%%%%%%%%%%%%%%%%%%%%%%%%%%%%%%%%%%%%%%%%%%%%%%
\subsection{Density matrix of emitted photon}
\label{ssec:dm}
%%%%%%%%%%%%%%%%%%%%%%%%%%%%%%%%%%%%%%%%%%%%%%%%%%%%%%%%%%%%%%%%%%%%%%%%%%%%%%%
In the $t\to\infty$ limit,
the initial dot excitation is completely transformed into a photon
propagating in the intended mode ($b$ field)
or in unintended directions ($c$ field).
In the following subsections, we analyze the photon emitted in the $b$ field.
It is fully characterized by its density matrix $\hat{\rho}(t)$.
In the real-space representation,
the matrix element $\rho(r,r',t)$ is given by
\begin{align}
\rho(r,r',t) &= \la \psi(t)|\tb^\dag_{r'}\tb_r|\psi(t)\ra.
\end{align} 
We make the following three comments regarding this quantity:
(i)~$\hat{\rho}(t)$ is Hermitian, namely, $\rho(r',r,t)=\rho^{\ast}(r,r',t)$. 
Therefore, we need consider only the $r<r'$ region in this subsection.
(ii)~As we will see later, $\rho(r,r',t)=\rho(r-t,r'-t)$ in the $t\to\infty$ limit.
This reflects the translational motion of the emitted photon.
(iii)~Tr$\hat{\rho}(t)=\int dr \rho(r,r,t)$ represents the probability
of finding the emitted photon in the $b$ field.
This is unity when $\gamma=0$.

Using eqs.~(\ref{eq:vec}) and (\ref{eq:FactorizeGamma}), 
the matrix element can be rewritten as
\begin{align}
\rho(r,r',t) 
&= \kappa \beta_0(t-r) \beta_0^{\ast}(t-r')
+2\gammap\kap\int dx \beta_0(x-r) \beta_0^{\ast}(x-r')|\alp_0(t-x)|^2
+\cdots.
\end{align}
We here introduce the Laplace transform of $\beta_0\beta_0^{\ast}$, which is
defined by $\cL_{\beta_0\beta_0^{\ast}}(r,r',z) 
= \int_0^{\infty} dt e^{-zt} \beta_0(t-r) \beta_0^{\ast}(t-r')$.
It is given by
\begin{align}
\cL_{\beta_0\beta_0^{\ast}}(r,r',z) 
&= \sum_{m,n =1,2} \frac{B_m B_n^{\ast}}{z-\lambda_m-\lambda_n^{\ast}} 
e^{\lambda_m(r'-r)-r'z},
\end{align}
where $\lam_{1,2}$ and $B_{1,2}$ are defined in \S\ref{sec:sv}.
The Laplace transform of $\rho(r,r',t)$ is then given by
\begin{align}
\cL_{\rho}(r,r',z) 
&= \frac{\kappa \cL_{\beta_0\beta_0^{\ast}}(r,r',z)}{1-2\gammap \cL_{|\alp_0|^2}(z)}.
\label{eq:cLrho}
\end{align}
By analyzing the poles of this function in the $z$-plane,
$\rho(r,r',t)$ is obtained as follows:
\begin{align}
\rho(r,r',t) 
&= \sum_{j=1}^{4} \sum_{m=1}^{2} \rho_{jm} 
e^{\lambda_m(r'-r)+\mu_j (t-r')},
\label{eq:rho1}
\\
\rho_{jm} &= \frac{\prod_{m', n'=1,2}(\mu_j-\lambda_{m'}-\lambda_{n'}^{\ast})}
{\prod_{i(\ne j)} (\mu_j - \mu_i)} 
\sum_{n=1}^{2} \frac{B_m B_n^{\ast}}{\mu_j - \lambda_m - \lambda_n^{\ast}},
\label{eq:rho2}
\end{align}
where $0<r<r'<t$ and $\mu_j$ ($j=1,\cdots,4$) are defined in \S\ref{ssec:decay}.
We can check that this quantity depends on only two variables, $r-t$ and $r'-t$.

%%%%%%%%%%%%%%%%%%%%%%%%%%%%%%%%%%%%%%%%%%%%%%%%%%%%%%%%%%%%%%%%%%%%%%%%%%%%%%%
\subsection{Pulse shape}
%%%%%%%%%%%%%%%%%%%%%%%%%%%%%%%%%%%%%%%%%%%%%%%%%%%%%%%%%%%%%%%%%%%%%%%%%%%%%%%
The pulse shape of the emitted photon is characterized 
by the intensity distribution
$f(r,t) = \la\psi(t)|\tb_r^{\dag}\tb_r|\psi(t)\ra = \rho(r,r,t)$,
which is the diagonal element of the density matrix.
This quantity is real and positive for $0<r<t$.
By setting $r'=r$ in eq.~(\ref{eq:rho1}), we have
\begin{align}
f(r,t) &= \sum_{j=1}^{4} f_j e^{\mu_j(t-r)}, \\
f_j &= \frac{\prod_{m^\prime, n^\prime = 1,2} 
(\mu_j - \lambda_{m^\prime} -\lambda_{n^\prime}^{\ast})}
{\prod_{i\ne j} (\mu_j - \mu_i)} \sum_{m,n=1}^{2} 
\frac{B_m B_n^{\ast}}{\mu_j - \lambda_m - \lambda_n^{\ast}}.
\end{align}

%%%%%%%%%%%%%%%%%%%%%%%%%%%%%%%%%%%%%%%%%%%%%%%%%%%%%%%%%%%%%%%%%%%%%%%%%%%%%%%
\subsection{Frequency spectrum}
%%%%%%%%%%%%%%%%%%%%%%%%%%%%%%%%%%%%%%%%%%%%%%%%%%%%%%%%%%%%%%%%%%%%%%%%%%%%%%%
The frequency spectrum of the emitted photon is defined by
$S(k,t)=\la\psi(t)|b_k^\dag b_k|\psi(t)\ra$.
Apparently, $S(k,t)$ becomes independent of $t$
in the $t\to\infty$ limit,
and we are interested in $S(k)=\lim_{t\to\infty}S(k,t)$.
By definition, $S(k,t)$ is the Fourier transform of 
the density matrix element:
\begin{align}
S(k,t)= \frac{1}{2\pi} \dint dr dr' e^{ik(r'-r)} \rho(r,r',t). 
\label{eq:Skt}
\end{align}
We consider the Laplace transform of $S(k,t)$
defined by $\cL_S(k,z)=\int_0^{\infty} dt e^{-zt}S(k,t)$.
Using eqs.~(\ref{eq:cLrho}) and (\ref{eq:Skt}), it is given by
\begin{align}
\cL_S(k,z)= \frac{\kap}{2\pi}
\frac{\cL'_{\beta_0\beta_0^{\ast}}(k,z)}{1-2\gammap\cL_{|\alp_0|^2}(z)}, 
\end{align}
where
\begin{align}
\cL'_{\beta_0\beta_0^{\ast}}(k,z) &= 
\int_0^{\infty} dt \dint dr dr' e^{ik(r'-r)-zt} 
\beta_0^{\ast}(t-r') \beta_0^{\ast}(t-r) \nonumber \\
&= \int_0^{\infty} dt e^{-zt} \left| \sum_{m=1}^{2} 
\frac{B_m}{\lambda_m + ik}(e^{-ikt} - e^{\lambda_m t}) \right|^2.
\end{align}
Note that $\cL'_{\beta_0\beta_0^{\ast}}$ has a pole at $z=0$.
Since our interest lies in the $t\to\infty$ limit of $S(k,t)$,
we must investigate the pole of $\cL_S(k,z)$ at $z=0$ only.
Therefore, $S(k)=(\kap/2\pi)[1-2\gammap\cL_{|\alp_0|^2}(0)]^{-1}
\Res_{z=0}[\cL'_{\beta_0\beta_0^{\ast}}(k,z)]$.
After some calculations, we obtain
\begin{align}
S(k) &= \frac{\cN}{|(k-\tom_d)(k-\tom_c)-g^2|^2},
\end{align}
where $\cN=(\kap g^2/2\pi)[1-2\gammap\cL_{|\alp_0|^2}(0)]^{-1}$
is a factor that is independent of $k$.
This spectral shape was predicted by Glauber~\cite{Glauber65},
and it was confirmed in recent theoretical studies
based on 
the quantum Langevin equations~\cite{Cui06,Naesby08}
and the Master equations~\cite{Yamaguchi08,Auffeves09,Auffeves10}.
The relation to the Master equation approach is summarized in Appendix \ref{APPPerturbation}.

%%%%%%%%%%%%%%%%%%%%%%%%%%%%%%%%%%%%%%%%%%%%%%%%%%%%%%%%%%%%%%%%%%%%%%%%%%%%%%%
\subsection{Purity}
\label{sec:Purity}
%%%%%%%%%%%%%%%%%%%%%%%%%%%%%%%%%%%%%%%%%%%%%%%%%%%%%%%%%%%%%%%%%%%%%%%%%%%%%%%
%-------------------------------------------------------------------------------
\begin{figure}[t]
\begin{center}
\includegraphics[width=0.4 \columnwidth]{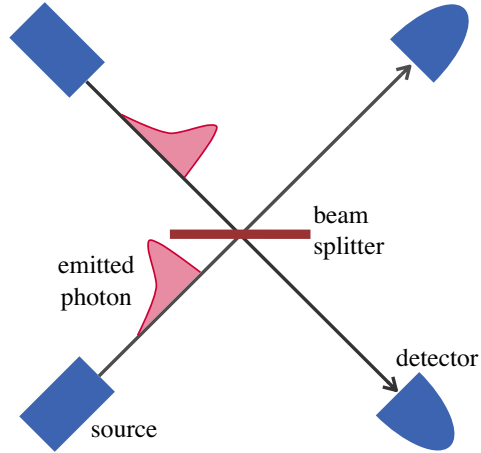}
\caption{(Color online) Schematic illustration of two-photon interference experiment.
The coincidence probability vanishes 
when the two input photons are completely indistinguishable.}
\label{fig:2pi}
\end{center}
\end{figure}
%-------------------------------------------------------------------------------

Quantum information processing requires
high indistinguishability between single photons.
A popular measure of the indistinguishability of photons is 
the coincidence probability $P_{co}$ in two-photon interference experiments
(see Fig.~\ref{fig:2pi}).
Two solid-state emitters simultaneously
emit single photons into two input ports of a beam splitter.
These two photons are mixed by the beam splitter
and are counted by photo detectors.. 
When two indistinguishable photons are simultaneously input to a beam splitter,
they always appear at the same output port
(Hong--Ou--Mandel interference), namely, $P_{co}=0$.
However, pure dephasing generates entanglement between an emitted photon 
and the environment of its source,
making two-photon interference imperfect ($P_{co}>0$).
The coincidence probability is related to the purity $\cP$ 
of a photon by $P_{co}=(1-\cP)/2$ (see Appendix~\ref{app:relation} for the derivation).

The purity is defined in terms of the density matrix $\hat{\rho}(t)$ 
by $\cP=\mathrm{Tr}[\hat{\rho}^2(t)]$.
As expected, this quantity becomes independent of $t$ 
when $t$ is sufficiently large.
Using the real-space matrix element, the purity can be rewritten as
\begin{align}
\mathcal{P} &= 
\lim_{t\rightarrow\infty}\int dr dr' \rho(r,r',t)\rho(r',r,t)
= \lim_{t\rightarrow\infty}\int dr dr' |\rho(r,r',t)|^2.
\end{align}
Using eq.~(\ref{eq:rho2}), we have
\begin{align}
\mathcal{P} &= \sum_{j,j'=1}^{4} \sum_{m,m'=1}^{2} 
\frac{2\rho_{jm} \rho_{j'm'}^*}
{(\mu_j + \mu_{j'}^*)(\lambda_m + \lambda_{m'}^*)}.
\end{align}

%%%%%%%%%%%%%%%%%%%%%%%%%%%%%%%%%%%%%%%%%%%%%%%%%%%%%%%%%%%%%%%%%%%%%%%%%%%%%%%
\section{Numerical Results}
\label{sec:num}
%%%%%%%%%%%%%%%%%%%%%%%%%%%%%%%%%%%%%%%%%%%%%%%%%%%%%%%%%%%%%%%%%%%%%%%%%%%%%%%
The analytical results derived in the previous section
are rigorous and applicable to any set of parameters,
$(\om_d-\om_c, g, \kap, \gamma, \gamma_p)$.
In this section, we visualize these results
by employing specific parameters.
Throughout this section, we assume, for simplicity, that 
radiative decay to unintended modes does not occur ($\gamma=0$).

%%%%%%%%%%%%%%%%%%%%%%%%%%%%%%%%%%%%%%%%%%%%%%%%%%%%%%%%%%%%%%%%%%%%%%%%%%%%%%%
\subsection{Zeno and anti-Zeno effects}
\label{ssec:zaz}
%%%%%%%%%%%%%%%%%%%%%%%%%%%%%%%%%%%%%%%%%%%%%%%%%%%%%%%%%%%%%%%%%%%%%%%%%%%%%%%
%-------------------------------------------------------------------------------
\begin{figure}[h]
\begin{center}
\includegraphics[width=0.9 \columnwidth]{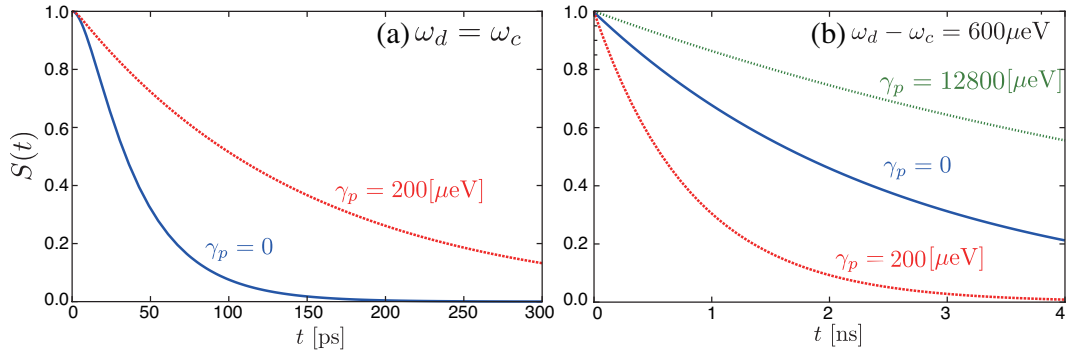}
\caption{(Color online) Survival probability $S(t)$ for 
(a)~resonant ($\om_d=\om_c$) and (b)~detuned ($\om_d-\om_c= 600$ $\mathrm{\mu eV}$)  cases.
$g = 25$ $\mathrm{\mu eV}$ and $\kappa = 150$ $\mathrm{\mu eV}$.
The values of $\gamma_p$ are indicated in the figures.}
\label{fig:sv}
\end{center}
\end{figure}
%-------------------------------------------------------------------------------
\begin{figure}[h]
\begin{center}
\includegraphics[width=0.55 \columnwidth]{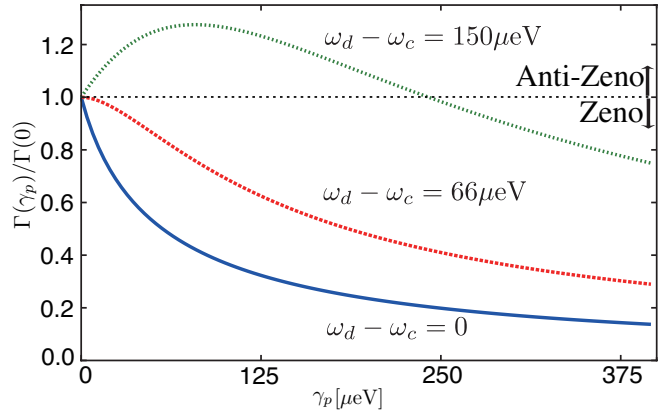}
\caption{(Color online) Dependence of the radiative decay rate $\tilde{\Gamma}$
on the pure dephasing rate $\gamma_p$.
The radiative decay rate is normalized 
by the {\it free} decay rate (i.e., the rate for $\gamma_p=0$).
$g = 25$ $\mathrm{\mu eV}$ and $\kappa = 150$ $\mathrm{\mu eV}$.
$\om_d-\om_c=0$ (solid), $66$ (dotted) 
and $150$ $\mathrm{\mu eV}$ (dashed).}
\label{fig:rate}
\end{center}
\end{figure}
%-------------------------------------------------------------------------------

First, we observe the effects of pure dephasing on the decay of dot excitation.
We focus on the weak-coupling regime ($\kap=6g$) in this subsection,
where the dot decays monotonically without revival
and obeys the exponential decay law with high accuracy.
The decay rate of the dot is well defined in this case
and is given by
$\Gamma=\lim_{t\rightarrow\infty}[-\log P(t)/t]$.
This reduces to $\min_j |{\rm Re}\ \mu_j|$,
where $\mu_j$ is defined in \S\ref{ssec:decay}.
Figure~\ref{fig:sv} shows the temporal behavior of the survival probability $P(t)$.
In Fig.~\ref{fig:sv}(a), 
where the dot is in resonance with the cavity ($\om_d=\om_c$),
the decay becomes slower as pure dephasing increases.
In contrast, in Fig.~\ref{fig:sv}(b), 
where the dot is detuned from the cavity ($\om_d-\om_c=600$ $\mathrm{\mu eV}$),
the decay becomes faster under small pure dephasing ($\gamma_p=200$ $\mathrm{\mu eV}$),
whereas the decay becomes slower under larger pure dephasing ($\gamma_p=12800$ $\mathrm{\mu eV}$).

Measurements to check the survival of dot excitation
destroy the quantum coherence between the excited and ground states
that preserves the population of these two states.
Therefore, pure dephasing has the same effect
on the dot as continuous measurements,
if the measurement results are unquestioned.
The changes in the decay rates induced by pure dephasing
are often interpreted as the quantum Zeno and anti-Zeno effects
\cite{Kofman01,Shnirman03,Yamaguchi08}.
Previous analyses indicated that the anti-Zeno effect can be observed
only when the dot--cavity detuning is large
and pure dephasing is small~\cite{Koshino05}.
This agrees with the present numerical results.
Figure~\ref{fig:rate} shows the dependence of the radiative decay rate
$\Gamma$
on the pure dephasing rate $\gamma_p$.
To clearly observe the Zeno and anti-Zeno effects,
$\Gamma(\gamma_p)$ is normalized by the {\it free} decay rate of
$\Gamma(0)$;
$\Gamma(\gamma_p)/\Gamma(0)<1$ indicates the Zeno effect,
whereas $\Gamma(\gamma_p)/\Gamma(0)>1$ exhibits the anti-Zeno effect.
Non monotonic behavior of $\Gamma(\gamma_p)$ is clearly observed for
large detuning.

%%%%%%%%%%%%%%%%%%%%%%%%%%%%%%%%%%%%%%%%%%%%%%%%%%%%%%%%%%%%%%%%%%%%%%%%%%%%%%%
\subsection{Pulse shape and spectrum}
\label{sec:SpectrumResult}
%%%%%%%%%%%%%%%%%%%%%%%%%%%%%%%%%%%%%%%%%%%%%%%%%%%%%%%%%%%%%%%%%%%%%%%%%%%%%%%
%-------------------------------------------------------------------------------
\begin{figure}[tb]
\begin{center}
\includegraphics[width=0.95 \columnwidth]{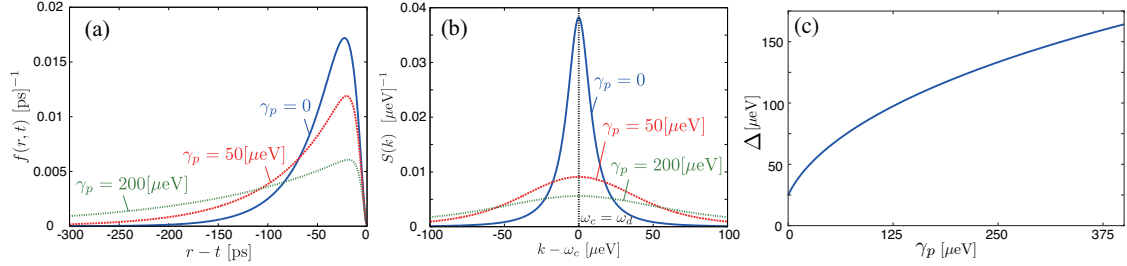}
\caption{(Color online) (a)~Pulse shape $f(r,t)$ and (b)~spectrum $S(k)$ of emitted photon.
$g = 25$ $\mathrm{\mu eV}$, $\kappa = 150$ $\mathrm{\mu eV}$, and $\om_d-\om_c=0$.
$\gamma_p=0$ (solid), $50$ (dotted), and $200$ $\mathrm{\mu eV}$ (dashed).
(c)~Spectral width as a function of $\gamma_p$.}
\label{fig:OnResonant}
\end{center}
\end{figure}
%-------------------------------------------------------------------------------
In this subsection, we examine the pulse shape 
and frequency spectrum of the emitted photon
using the same parameters as those used in the previous subsection.
First, we observe the results for the resonant ($\om_d=\om_c$) case.
The pulse shapes $f(r,t)$ of the emitted photon are shown in Fig.~\ref{fig:OnResonant}(a)
for three pure dephasing rates  $\gamma_p$.
Each pulse shape is normalized 
[$\int_{-\infty}^t dr f(r,t) = 1$]
since $\gamma=0$ is assumed here
and a single photon is necessarily generated in the $b$ field.
The pulse becomes longer as pure dephasing increases.
This is consistent with the quantum Zeno effect 
discussed in the previous subsection:
in the resonant case, the decay of the dot 
becomes monotonically slower with increasing pure dephasing.
Figure~\ref{fig:OnResonant}(b) shows the spectra $S(k)$ of the photon
for the same parameters as those in Fig.~\ref{fig:OnResonant}(a).
These spectra have a single peak at $k=\om_d(=\om_c)$ 
and are normalized [$\int_{-\infty}^{\infty} dk S(k) = 1$].
The spectrum broadens with increasing pure dephasing.
This is confirmed by Fig.~\ref{fig:OnResonant}(c),
in which the spectral width
(defined as $\Delta=[\int dk (k-\om_d)^2 S(k)]^{1/2}$)
is plotted as a function of the pure dephasing rate.
Thus, 
the pulse broadens in both the real and frequency spaces with increasing $\gamma_p$
and it is thus not Fourier-limited.
This implies that the emitted photon is in a mixed state when $\gamma_p\neq 0$.
The purity of the photon is discussed later in \S\ref{ssec:pur}.

%-------------------------------------------------------------------------------
\begin{figure}[tb]
\begin{center}
\includegraphics[width=0.9 \columnwidth]{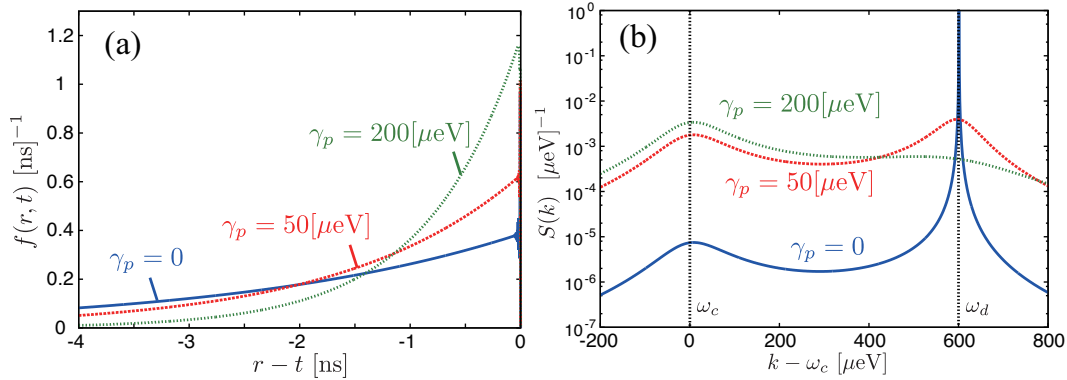}
\caption{(Color online) (a)~Pulse shape $f(r,t)$ and (b)~spectrum $S(k)$ of emitted photon.
$g = 25$ $\mathrm{\mu eV}$, $\kappa = 150$ $\mathrm{\mu eV}$, and $\om_d-\om_c=600$ $\mathrm{\mu eV}$.
$\gamma_p=0$ (solid), $50$ (dotted), and $200$ $\mathrm{\mu eV}$ (dashed).}
\label{fig:detuned}
\end{center}
\end{figure}
%-------------------------------------------------------------------------------

Next, we observe the results for the detuned case.
Figure~\ref{fig:detuned}(a) shows the pulse shapes of the photon.
The pulse shape is approximately exponential,
except for the oscillatory behavior at the very initial stage.
The pulse length is inversely proportional to 
the decay rate of the dot.
Figure~\ref{fig:detuned}(b) shows the photon spectra 
for the same parameters as those in Fig.~\ref{fig:detuned}(a).
A notable difference from the resonant case is that
the spectra are doubly peaked with peaks at both the dot frequency $\om_d$
and the cavity frequency $\om_c$. 
The widths of these peaks are determined by
$\left|\mathrm{Im}\tom_d\right|=\gamma/2+\gamma_p$ and 
$\left|\mathrm{Im}\tom_c\right|=\kappa/2$.
Therefore, the width of the peak at $\om_d$ is sensitive to pure dephasing.
When pure dephasing is weak ($\gamma_p\ll\kap$),
as in atomic cavity QED systems,
the dominant peak of $S(k)$ appears at $\om_d$.
In contrast, when pure dephasing is strong 
($\gamma_p\gg\kap$), as in solid-state systems,
the dominant peak of $S(k)$ appears at $\om_c$.
This behavior can be seen more clearly in an approximate form of the spectrum,
\begin{eqnarray}
S(k) &=& \frac{2\gamma_p}{2\gamma_p+\kappa} S_{\rm cav}(k) + 
\frac{\kappa}{2\gamma_p+\kappa} S_{\rm dot}(k), \label{eq:specdiv} \\
S_{\rm cav}(k) &=& \frac{\kappa/2\pi}{(k-\omega_c)^2+(\kappa/2)^2}, \\
S_{\rm dot}(k) &=& \frac{\gamma_p/\pi}{(k-\omega_d)^2+\gamma_p^2},
\end{eqnarray}
which is valid for $|\omega_d - \omega_c| \gg g$. From this expression,
it is proved that 
the cavity spectrum $S_{\rm cav}$ is dominant for $\gamma_p\gg\kappa$,
whereas the dot spectrum $S_{\rm dot}$ is dominant for $\gamma_p\ll\kappa$.
This result partly explains the detuned peaks observed
in the resonance fluorescence spectrum in solid-state cavity QED systems,
as discussed in previous theoretical works~\cite{Cui06,Naesby08,Yamaguchi08}.

The mean energy of the emitted photon is evaluated by
$E_p=\int dk \ kS(k)$.
Owing to energy conservation,
the mean photon energy is expected to always be identical to the dot frequency
(i.e., $E_p=\om_d$).
However, Fig.~\ref{fig:detuned}(b) clearly shows that the mean photon energy
is sensitive to pure dephasing
and may deviate from the dot frequency when pure dephasing is present. 
This discrepancy can be resolved by considering the energy 
released to the environment during decay,
which is given by $E_e=\int dk \ k\la d_k^{\dagger}d_k\ra$.
It can be shown that (see Appendix~\ref{APPEnergy} for derivation)
\begin{align}
E_p &= 
\frac{2\gammap}{\kappa+2\gammap} \omega_c +
\frac{\kappa}{\kappa+2\gammap} \omega_d,
\label{EnergyE} \\
E_e &= \frac{2\gammap}{\kappa+2\gammap}(\om_d-\om_c).
\label{EnergyB}
\end{align}
Thus, energy conservation is satisfied 
when the environmental energy is included ($E_p+E_e=\om_d$).
It should be noted that
while pure dephasing coupling 
never induces a transition in the dot,
this coupling enables energy exchange
between the dot and the environment.
When the cavity frequency exceeds the dot frequency,
eq.~(\ref{EnergyB}) indicates that the dot may absorb 
environmental energy during decay~\cite{recentPRA}.

%%%%%%%%%%%%%%%%%%%%%%%%%%%%%%%%%%%%%%%%%%%%%%%%%%%%%%%%%%%%%%%%%%%%%%%%%%%%%%%
\subsection{Purity}
\label{ssec:pur}
%%%%%%%%%%%%%%%%%%%%%%%%%%%%%%%%%%%%%%%%%%%%%%%%%%%%%%%%%%%%%%%%%%%%%%%%%%%%%%%

As discussed in \S\ref{sec:Purity},
the coincidence probability in two-photon interference $P_{co}$,
which is a popular measure of indistinguishability,
is related to the purity as $P_{co}=(1-\cP)/2$.
In this subsection, we show numerical results for the purity $\cP$ for various parameter regions. 
Qualitative features in some limiting cases are discussed in Appendix~\ref{app:pur_perturbation} on the basis of perturbation calculation.

%-------------------------------------------------------------------------------
\begin{figure}[tb]
\begin{center}
\includegraphics[width=0.7 \columnwidth]{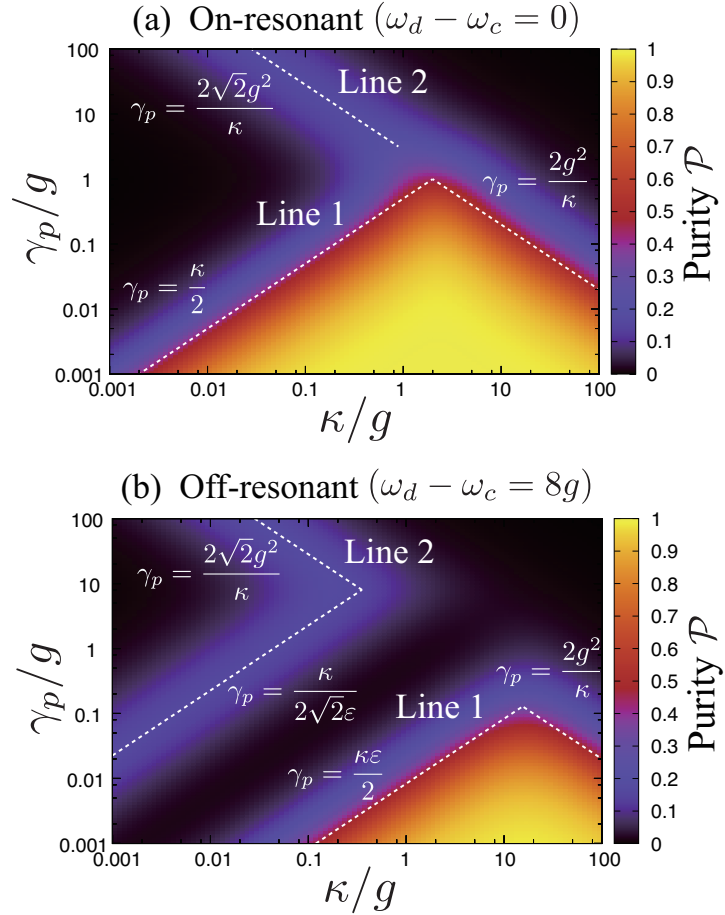}
\caption{(Color online) Purity calculated as a function of $\kappa/g$ and $\gamma_p/g$
for (a) the resonant case ($\omega_d=\omega_c$) and (b) the detuned case ($\omega_d-\omega_c=8g$).}
\label{fig:Purity}
\end{center}
\end{figure}
%-------------------------------------------------------------------------------

We first consider the resonant case ($\omega_c = \omega_d$).
In Fig. \ref{fig:Purity}(a), the purity $\cP$ is plotted as a function of $\kappa$ and $\gamma_p$.
The purity becomes large for $\gamma_p < \min(\kappa/2, 2g^2/\kappa)$
(the region below `Line 1' in the figure).
From the figure, we can see that 
for the generation of coherent photons, pure dephasing should be reduced
as much as possible, and that for fixed pure dephasing,
$\kappa$ should be taken near the point of critical damping ($\kappa = 2g$). 
In addition to this main region appropriate for highly coherent photon generation,
there is a line on which the purity has a small peak (denoted as `Line 2' in the figure)
in the region of $\gammap \gg g$. 
By perturbation calculation (see Appendix~\ref{app:pur_perturbation}),
we can show that the purity takes a maximum value of $\cP_{\rm max} = 3-2\sqrt{2} \simeq 0.17$ 
at $\gamma_p = 2\sqrt{2}g^2/\kappa$.

Next, we consider the strongly detuned case ($|\omega_d - \omega_c| \gg g$).
In Fig. \ref{fig:Purity}(b), the purity is shown as a function of $\kappa$ and $\gamma_p$
for $\omega_d - \omega_c = 8g$.
The purity has a large value below the line $\gamma_p < \min(\varepsilon \kappa/2, 2g^2/\kappa)$
(the region below `Line 1' in the figure), where $\varepsilon = g^2/(\omega_d-\omega_c)^2$
is a small parameter dependent on detuning. In Fig. \ref{fig:Purity}(b), we can see
that the purity for the detuned case is suppressed compared with the resonant case, and that the high-purity region is narrowed.
From the figure, we can see that
for coherent photon emission, dephasing should be reduced as much as possible, and
that for fixed pure dephasing, $\kappa$ should be taken to be around $\kappa = (2\sqrt{2}g^2/\varepsilon)^{1/2}$.
In addition to the main high-purity region, the purity has a small peak (denoted as `Line 2') similarly to the resonant case.
By perturbation calculation, we can show that the purity has a maximum value of
$\cP_{\rm max} = 3-2\sqrt{2} \simeq 0.17$ at $\gamma_p = \kappa/(2\sqrt{2} \varepsilon),
2\sqrt{2}g^2/\kappa$.

\subsection{Time filtering}
\label{ssec:timefiltering}

%-------------------------------------------------------------------------------
\begin{figure}[tb]
\begin{center}
\includegraphics[width=0.9 \columnwidth]{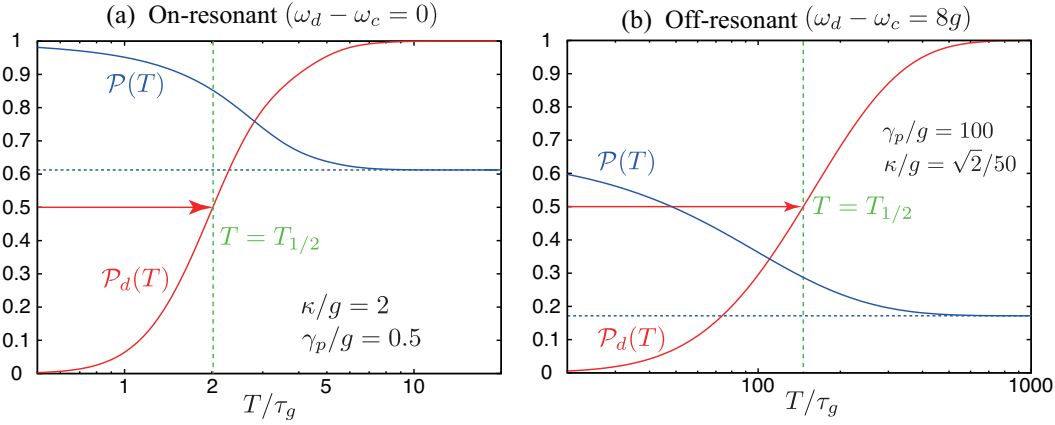}
\caption{(Color online) Effective purity $\cP(T)$ and efficiency $\cP_d(T)$ 
of photons emitted to the output port during $0 < t <T$ shown
as a function of $T$. (a) On-resonant case ($\omega_d = \omega_c$)
for $\kappa/g=2$ and $\gamma_p/g = 0.5$, and (b) detuned case ($\omega_d
- \omega_c = 8g$) for $\kappa/g = \sqrt{2}/50$ and $\gamma_p/g = 100$.
The time $T_{1/2}$ giving half-efficiency $\cP_d(T_{1/2})=1/2$ is also shown
by a green line. The unit of time is given as $\tau_g = \hbar/g$ and is about $25$ ${\rm ps}$                                                                                                                                                                                                                                                                                                                                                                                                                                                                                                                                                                                                                                                                                                                                                                                                                                                                                                                                                                                                                                                                                                                                                                                                                                                                                                                                                                                                   for $g = 25$ $\mu{\rm eV}$.
}
\label{fig:Timefilter}
\end{center}
\end{figure}
%-------------------------------------------------------------------------------

A strategy for improving the purity of emitted photons is to use
photons emitted at the early stage of decay, 
because such photons are subject to the environment only for a short time.
The density matrix of photons filtered in the temporal region $0 \le t - r \le T$ is written 
as $\rho_T(r,r',t) = \rho(r,r',t) \Theta(T-t+r) \Theta(T-t+r')$. From $\rho_T(r,r',t)$,
the purity and efficiency of filtered photons are given as
\begin{eqnarray}
\mathcal{P}(T) &=&
\frac{{\rm Tr}(\rho_T^2)}{\mathcal{P}_d(T)} = \frac{1}{\mathcal{P}_d}\int_{t-T}^{t} dr \int_{t-T}^{t} dr' \
|\rho(r,r',t)|^2, \\
\mathcal{P}_d(T) &=& \left[{\rm Tr}(\rho_T)\right]^2 = \left(\int_{t-T}^{t} dr \ \rho(r,r,t) \right)^2,
\end{eqnarray}
where $\mathcal{P}_d(T)$ is the square of the probability that a single photon is emitted
during $0 \le t \le T$.
$\mathcal{P}_d(T)$ is also the square of the probability that a single photon is obtained after time filtering.
By time filtering,
the effective purity $\mathcal{P} (T)$ is improved at the expense of
the square of the probability $\mathcal{P}_d(T)$.

We first discuss the resonant ($\omega_d=\omega_c$) case.
In Fig.~\ref{fig:Timefilter}(a), we show the $T$-dependence of $\mathcal{P}(T)$ and $\mathcal{P}_d(T)$
for $\kappa/g = 2$ and $\gamma_p/g = 0.5$. 
In the $T\rightarrow\infty$ limit,
$\mathcal{P}(T)$ approaches 0.61, which is the purity without time filtering.
As $T$ is shortened, the purity $\mathcal{P}(T)$ increases, whereas the square of the probability 
$\mathcal{P}_d(T)$ decreases.
If we allow the reduction of the square of the probability to $1/2$, one can set the filtering time
at $T=2\tau_g$ [see the vertical dashed line in Fig.~\ref{fig:Timefilter}(a)],
and obtain an improved purity of $0.85$.
Improvement of the purity is also possible in the detuned case.
In Fig.~\ref{fig:Timefilter}(b), we show the $T$-dependence of $\mathcal{P}(T)$ and $\mathcal{P}_d(T)$
for $\kappa/g = \sqrt{2}/50$, $\gamma_p/g = 100$, and $\omega_d - \omega_c = 8g$, 
which is on the upper line of `Line 2' in Fig.~\ref{fig:Purity}(b). 
If we allow the reduction of the square of the probability to $1/2$, the purity is improved from 0.17 to 0.28.

We can improve the purity further by choosing a shorter
filtering time. The efficiency, however, decreases exponentially in the limit $T\rightarrow 0$.

%%%%%%%%%%%%%%%%%%%%%%%%%%%%%%%%%%%%%%%%%%%%%%%%%%%%%%%%%%%%%%%%%%%%%%%%%%%%%%%
\section{Summary} 
%%%%%%%%%%%%%%%%%%%%%%%%%%%%%%%%%%%%%%%%%%%%%%%%%%%%%%%%%%%%%%%%%%%%%%%%%%%%%%%
We analyzed the radiative decay of an excited dot in a solid-state cavity QED system
and investigated the quantum-mechanical properties of the emitted photon,
stressing the effects of pure dephasing.
Our analysis is based on a model in which all the elements of the system
(including environmental ones) are treated 
as active quantum-mechanical degrees of freedom.
We rigorously solved the time evolution of the overall system
and derived analytical expressions for the density matrix, pulse shape, 
spectrum, and purity of the emitted photon.
These analytical results were visualized under realistic parameters.
The main results are summarized as follows.
(i)~Changes in the dot decay rate owing to pure dephasing
can be explained in terms of the quantum Zeno and anti-Zeno effects.
The emitted photon pulse length is approximately given 
by the inverse of the dot decay rate.
(ii)~The emitted photon spectrum agrees with the Glauber formula.
The mean energy of an emitted photon is not necessarily identical to that of the dot
and energy conservation is seemingly broken.
However, the present analysis revealed that 
the dot can exchange energy with the environment through pure dephasing coupling.
Energy conservation holds when the environmental energy is included.
(iii)~The purity of the emitted photon is calculated 
as a measure of indistinguishability. 
To generate pure single photons, dephasing should be reduced as much as possible
and the escape rate of cavity photons $\kappa$ should be taken near the optimal value ($\kappa=2g$ for the resonant case and $\kappa=(2\sqrt{2}g^2/\epsilon)^{1/2}$ for the strongly detuned case).
(iv)~Time filtering improves the purity of the emitted photon at the expense of a lower probability of single photon generation.

In our approach, we have assumed white noise for the level fluctuation of the QD. 
For a detailed comparison with results of experiments, we must consider more realistic models taking into account 
phonons~\cite{Tarel10,Kaer10} and background carriers~\cite{Yamaguchi08}. 
The extension of the present approach to these extended models is an important problem left for future study.

%%%%%%%%%%%%%%%%%%%%%%%%%%%%%%%%%%%%%%%%%%%%%%%%%%%%%%%%%%%%%%%%%%%%%%%%%%%%%%%
\section*{Acknowledgments}
%%%%%%%%%%%%%%%%%%%%%%%%%%%%%%%%%%%%%%%%%%%%%%%%%%%%%%%%%%%%%%%%%%%%%%%%%%%%%%%
This research was partially supported by 
MEXT KAKENHI (Grant Nos. 21740220, 22244035, and 23104710),
National Institute of Information and Communication Technology (NICT),
and the Strategic Information and Communications R \& D 
Promotion Program (SCOPE No.~111507004) of the
Ministry of Internal Affairs and Communications of Japan.

\appendix
%%%%%%%%%%%%%%%%%%%%%%%%%%%%%%%%%%%%%%%%%%%%%%%%%%%%%%%%%%%%%%%%%%%%%%%%%%%%%%%
\section{Relation to the Master Equation and Perturbation Method}
%%%%%%%%%%%%%%%%%%%%%%%%%%%%%%%%%%%%%%%%%%%%%%%%%%%%%%%%%%%%%%%%%%%%%%%%%%%%%%%
\label{APPPerturbation}

To analyze the properties of emitted photons, one can employ
the Master equation approach as an alternative method.
The Master equation is derived for the present model as
\begin{equation}
\frac{d}{dt} \left( \begin{array}{c} 
\rho_{\sigma \sigma} \\
\rho_{\sigma a} \\
\rho_{a \sigma} \\
\rho_{aa} \end{array} \right)
= - \left( \begin{array}{cccc} \gamma & ig & -ig & 0 \\
ig & \frac{\gamma + \kappa}{2} + \gamma_p + i \Delta \omega & 0 & -ig \\
-ig & 0 & \frac{\gamma + \kappa}{2} + \gamma_p - i\Delta \omega & ig \\
0 & -ig & ig & \kappa
\end{array} \right)
 \left( \begin{array}{c} 
\rho_{\sigma \sigma} \\
\rho_{\sigma a} \\
\rho_{a \sigma} \\
\rho_{aa} \end{array} \right),
\label{eq:mastereq}
\end{equation}
where $\rho_{\sigma \sigma} = \langle \sigma^{\dagger} \sigma \rangle$,
$\rho_{\sigma a} = \langle a^{\dagger} \sigma \rangle$,
$\rho_{a \sigma} = \langle \sigma^{\dagger} a \rangle$,
$\rho_{aa} = \langle a^{\dagger} a \rangle$, and
$\Delta \omega = \omega_d-\omega_c$. 
Using the solution of the master equation, the density matrix $\rho(r,r',t)$
of emitted photons can be written as
\begin{equation}
\rho(r,r',t) = \left\{ \begin{array}{ll} 
\rho_{\sigma a} (t-r') \beta_0(r'-r) + \rho_{aa} (t-r') \tilde{\beta}_0(r'-r)
& (r<r'<t), \\
\rho_{a \sigma} (t-r) \beta_0(r-r') + \rho_{aa} (t-r) \tilde{\beta}_0(r-r')
& (r'<r<t). \end{array} \right.
\end{equation}
Here, we have defined $(\tilde{\alpha}_0(t),\tilde{\beta}_0(t))$
as the solutions of eq.~(\ref{eq:a0b0}) under the initial condition $(\tilde{\alpha}_0(0),
\tilde{\beta}_0(0))=(0,1)$.

The four eigenvalues of the matrix in eq.~(\ref{eq:mastereq}) correspond to the values of $\mu_i$.
Therefore, numerical calculation of the eigenvalues of this matrix is a convenient 
method for obtaining the values of $\mu_i$. It is also useful to treat the matrix in eq.~(\ref{eq:mastereq}) 
in the perturbation calculation given in Appendix \ref{app:pur_perturbation}.
Because the matrix in eq.~(\ref{eq:mastereq}) is a normal matrix, it is diagonalized
by a unitary matrix. Therefore, the perturbation method used
in quantum mechanics can be applied to obtain the small shift of the values of $\mu_i$ with respect
to small parameters.

%%%%%%%%%%%%%%%%%%%%%%%%%%%%%%%%%%%%%%%%%%%%%%%%%%%%%%%%%%%%%%%%%%%%%%%%%%%%%%%
\section{Expressions for Purity in Limiting Cases}
%%%%%%%%%%%%%%%%%%%%%%%%%%%%%%%%%%%%%%%%%%%%%%%%%%%%%%%%%%%%%%%%%%%%%%%%%%%%%%%
\label{app:pur_perturbation}

A simple expression for $\mathcal{P}$ can be obtained in some limiting cases.
We first discuss the strongly detuned case ($|\omega_d - \omega_c| \gg g$).
The purity can then be expressed by a sum of two contributions 
as $\cP = \cP_{\rm dot} + \cP_{\rm cav}$, where $\cP_{\rm dot}$ and $\cP_{\rm cav}$ are the purities 
of photons whose frequencies are $\omega_d$ and $\omega_c$, respectively~\cite{footnote}. The former contribution
$\cP_{\rm dot}$ becomes large only below `Line 1' in Fig. \ref{fig:Purity}(b),
whereas $\cP_{\rm cav}$ has a maximum on `Line 2' in the figure. 
For $\gamma_p, \kappa \ll g$, perturbation calculation gives
\begin{equation}
\cP_{\rm dot} = \frac{\kappa}{2\gamma_p+\kappa} \times \frac{\kappa \varepsilon}{2\gamma_p + \kappa \varepsilon}.
\label{eq:approximate5}
\end{equation}
The first factor is the weight of the spectrum peak at $k = \omega_d$ [see eq.~(\ref{eq:specdiv})], and
the second one is that due to spectrum broadening by pure dephasing. For large detuning ($\varepsilon \ll 1$),
$\cP_{\rm dot}$ takes a value of 1/2 around $\gamma_p = \varepsilon \kappa/2$, 
which corresponds to the left line of `Line 1' in Fig.~\ref{fig:Purity}(b).
For $\gamma_p \ll g \ll \kappa$, perturbation calculation with respect to $g$ is effective. 
Then, one can consider the effective continuum constructed by mixing between the cavity mode 
and the output mode ($b$ field), and one can define the photon emission rate $\Gamma$ from the dot
into the effective continuum.
By using Fermi's golden rule, the photoemission rate is estimated as $\Gamma = 2\pi g^2 D(\omega_d)$,
where $D(\omega_d) = (\kappa/\pi)/((\omega_d - \omega_c)^2+(\kappa/2)^2)$ is the density of states
of the effective continuum. In the presence of pure dephasing,
the purity of photons emitted from a dot with the rate $\Gamma$ is calculated 
as $\cP_{dot} = 1/(1+2\gamma_p/\Gamma)$. Thus, we finally obtain
\begin{equation}
\cP_{\rm dot} = \frac{2g^2}{2g^2 + \gamma_p \kappa}
\end{equation}
for $\kappa \gg |\omega_d - \omega_c|$. One can see that the purity takes a value of 1/2 at
$\gamma_p = 2g^2/\kappa$, which corresponds to the right line of `Line 1' in Fig.~\ref{fig:Purity}(b).
For $\gamma_p, \kappa \ll g$, perturbation calculation of $\cP_{\rm can}$ gives
\begin{equation}
\cP_{\rm cav} = \frac{2\gamma_p \varepsilon \kappa}{(\kappa + 2\gamma_p \varepsilon)
(\kappa + 4\gamma_p \varepsilon)}.
\end{equation}
From this expression, it is proved that the purity takes a maximum value of $3-2\sqrt{2}$
at $\gamma_p =  \kappa/(2\sqrt{2} \varepsilon)$, which corresponds to the lower line
of `Line 2' in Fig.~\ref{fig:Purity}(b). For $\kappa \ll g \ll \gamma_p$, 
the dynamics of the cavity-dot system becomes completely incoherent and can be described
by the rate equations
\begin{eqnarray}
\dot{\rho}_{\sigma \sigma}(t) &=& - \Gamma' \rho_{\sigma \sigma}(t) + \Gamma' \rho_{aa}(t), \\
\dot{\rho}_{aa} (t) &=& +\Gamma' \rho_{\sigma \sigma}(t) - \Gamma' \rho_{aa}(t) - \kappa \rho_{aa}(t),
\end{eqnarray}
where $\rho_{\sigma \sigma}(t) = \langle \sigma^{\dagger}(t) \sigma(t) \rangle$,
$\rho_{aa}(t) = \langle a^{\dagger}(t) a(t) \rangle$, and $\Gamma' = 2 g^2/ \gamma_p$ is the transition
rate calculated from Fermi's golden rule. By solving these rate equations, we obtain
\begin{equation}
\cP_{\rm cav} = \frac{\kappa \Gamma'}{(\kappa + 2\Gamma')(\kappa + \Gamma')}.
\end{equation}
From this expression, it is proved that the purity takes a maximum value of $3-2\sqrt{2}$
at $\gamma_p = 2\sqrt{2} g^2/\kappa$, which corresponds to the upper line
of `Line 2' in Fig.~\ref{fig:Purity}(b). 

Next, we consider the resonant case ($\omega_c = \omega_d$). Some of the features
are common to the strongly detuned case;
the purity is reduced for $\kappa \ge 2g^2/\gamma_p$ [the right-side line of `Line 1'
in Fig.~\ref{fig:Purity}(a)] and takes a maximum value of $3-\sqrt{2}$ at $\gamma_p=
2\sqrt{2}g^2/\kappa$ [`Line 2' in Fig.~\ref{fig:Purity}(a)]. 
For $\gamma_p, \kappa \ll g$, the purity $\cP$ is approximately calculated as
\begin{equation}
\cP = \frac{\kappa(2\kappa+\gamma_p)}
{2(\kappa+\gamma_p)(\kappa + 2\gamma_p)}.
\end{equation}
From this expression, it is proved that $\cP \ge 1/2$ is realized
for the condition $\kappa \ge (1+\sqrt{3}) \gamma_p \simeq 2.73 \gamma_p$, which 
corresponds to the left line of `Line 1' in Fig.~\ref{fig:Purity}(a).

%%%%%%%%%%%%%%%%%%%%%%%%%%%%%%%%%%%%%%%%%%%%%%%%%%%%%%%%%%%%%%%%%%%%%%%%%%%%%%%
\section{Relation between Coincidence Probability and Purity}
\label{app:relation}
%%%%%%%%%%%%%%%%%%%%%%%%%%%%%%%%%%%%%%%%%%%%%%%%%%%%%%%%%%%%%%%%%%%%%%%%%%%%%%%
Here, we derive the relation $P_{co}=(1-\cP)/2$
between the coincidence probability $P_{co}$ and the purity $\cP$.
We consider the following situation (see Fig.~\ref{fig:2pi}).
Two solid-state emitters (S$_1$ and S$_2$) simultaneously and deterministically
emit single photons (assuming $\gamma=0$ for simplicity) 
into two input ports (P$_1$ and P$_2$) of a beam splitter.
These two photons are mixed by the beam splitter 
and are output to ports P$_3$ and P$_4$.
We denote the photon field operators for port P$_j$ 
by $\tb_{(j)r}$ ($j=1,\cdots,4$),
and the pure dephasing reservoir operators for emitter S$_j$
by $\td_{(j)r}$ ($j=1, 2$).
Assuming $t\to\infty$ and $\gamma=0$ (and thus $\delta_m=0$) in eq.~(\ref{eq:vec}),
the state vector of the photon at P$_1$ is given by
\begin{align}
|\psi_1\ra &=
\sum_{m=0}^{\infty}\int drd^m{\bm x}
\gamma_m(t,r,{\bm x}) 
\tb_{(1)r}^{\dag}|{\bm x}\ra,
\end{align}
where $|{\bm x}\ra=\td_{(1)x_1}^{\dag}\cdots\td_{(1)x_m}^{\dag}|0\ra$.
Thus, the emitted photon is entangled with the environment of its source.
The input state vector including the photons at both P$_1$ and P$_2$
is then given by
\begin{align}
|\psi_{12}\ra &=
\sum_{m,n=0}^{\infty}\int dr dr' d^m{\bm x} d^n{\bm x}' 
\ \gamma_m(t,r,{\bm x}) \gamma_{n}(t,r',{\bm x}') 
\tb_{(1)r}^{\dag}\tb_{(2)r'}^{\dag} |{\bm x};{\bm x}'\ra,
\end{align}
where $|{\bm x};{\bm x}'\ra=\td_{(1)x_1}^{\dag}\cdots\td_{(1)x_m}^{\dag}
\td_{(2)x'_1}^{\dag}\cdots\td_{(2)x'_n}^{\dag}|0\ra$.

The beam splitter mixes the two photons as
$\tb_{(1)r}^{\dag} \to [\tb_{(3)r}^{\dag} + \tb_{(4)r}^{\dag}]/\sqrt{2}$
and 
$\tb_{(2)r}^{\dag} \to [\tb_{(3)r}^{\dag} - \tb_{(4)r}^{\dag}]/\sqrt{2}$,
but it obviously does not affect the environmental degrees of freedom.
The output state vector is then given by
$|\psi_{out}\ra=|\psi_{33}\ra+|\psi_{44}\ra+|\psi_{34}\ra$,
where 
\begin{align}
& | \psi_{33}\ra = 
\sum_{m,n=0}^{\infty} \int dr dr' d^m{\bm x} d^n{\bm x}'
\frac{\g_m(t,r,{\bm x})\g_{n}(t,r',{\bm x}')}{2}
\tb_{(3)r}^{\dag}\tb_{(3)r'}^{\dag}  |{\bm x}; {\bm x}'\ra,  
% \nonumber 
\\
& |\psi_{44}\ra = 
- \sum_{m,n=0}^{\infty} \int dr dr' d^m{\bm x} d^n{\bm x}'
\frac{\g_m(t,r,{\bm x}) \g_n(t,r',{\bm x}')}{2}
\tb_{(4)r}^{\dag} \tb_{(4)r'}^{\dag}|{\bm x};{\bm x}'\ra,
% \nonumber 
\\
& | \psi_{34} \ra =
\sum_{m,n=0}^{\infty} \int dr dr' d^m{\bm x} d^n{\bm x}'
\frac{\g_m(t,r,{\bm x})\g_n (t,r',{\bm x}') - \g_m(t,r',{\bm x})\g_n(t,r,{\bm x}')}{2}
\tb_{(3)r}^{\dag}\tb_{(4)r'}^{\dag} |{\bm x}; {\bm x}'\ra. 
\label{eq:psi34}
\end{align}
It is readily confirmed that 
$\la\psi_{33}|\psi_{33}\ra+\la\psi_{44}|\psi_{44}\ra+\la\psi_{34}|\psi_{34}\ra=1$
and $\la\psi_{33}|\psi_{33}\ra=\la\psi_{44}|\psi_{44}\ra$.
The coincidence probability $P_{co}$
(i.e., the probability of finding single photons at both P$_3$ and P$_4$),
is given by $P_{co}=\la\psi_{34}|\psi_{34}\ra$.
Since the density matrix element is given by
$\rho(r,r',t)=\sum_{m=0}^{\infty}\int d^m{\bm x}
\gamma_m^{\ast}(r',{\bm x},t)\gamma_m(r,{\bm x},t)$,
$P_{co}$ is recast in the following form:
\begin{align}
P_{11}
&= \frac{1}{2}-\frac{1}{2}\int drdr' \rho(r,r',t)\rho(r',r,t) = \frac{1-{\cal P}}{2}.
\end{align}
When pure dephasing is present,
the purity of the emitted photon will be less than unity
and the coincidence probability will be nonzero.

%%%%%%%%%%%%%%%%%%%%%%%%%%%%%%%%%%%%%%%%%%%%%%%%%%%%%%%%%%%%%%%%%%%%%%%%%%%%%%%
\section{Proof of eqs.~(\ref{EnergyE}) and (\ref{EnergyB})}
\label{APPEnergy}
%%%%%%%%%%%%%%%%%%%%%%%%%%%%%%%%%%%%%%%%%%%%%%%%%%%%%%%%%%%%%%%%%%%%%%%%%%%%%%%
Here, we analytically evaluate $E_p=\int dk \ k \la b_k^\dag(t) b_k(t)\ra_i$
and $E_e=\int dk \ k \la d_k^\dag(t) d_k(t)\ra_i$ in the $t\to\infty$ limit,
where $\la\cdots\ra_i=\la\psi_i|\cdots|\psi_i\ra$.
Switching to real-space representations, we have
\begin{align}
E_p &= \frac{i}{2}\int dr 
\left\la(\partial_r\tb_r^{\dag})\tb_r-\tb_r^{\dag}(\partial_r\tb_r)\right\ra_i,
\\
E_e &= \frac{i}{2}\int dr 
\left\la(\partial_r\td_r^{\dag})\td_r-\td_r^{\dag}(\partial_r\td_r)\right\ra_i.
\end{align}
From eq.~(\ref{InputOutput1}), we obtain 
$\la(\partial_r b_r^{\dag})b_r\ra_i=-\kap\theta(t-\tau)
\theta(\tau)\la(\frac{d}{d\tau}a^{\dag})a\ra_i$,
where $\tau=t-r$. Using similar equations, we have
\begin{align}
E_p &= -\frac{i\kap}{2}\int_0^{\infty} d\tau 
\left\la({\textstyle \frac{d}{d\tau}}a^{\dag})a-a^{\dag}({\textstyle \frac{d}{d\tau}}a)\right\ra_i,
\\
E_e &= -i\gamma_p \int_0^{\infty} d\tau 
\left\la({\textstyle \frac{d}{d\tau}}\s^{\dag}\s)\s^{\dag}\s
-\s^{\dag}\s({\textstyle \frac{d}{d\tau}}\s^{\dag}\s)\right\ra_i.
\end{align}
Using eqs.~(\ref{HE1}) and (\ref{HE2}) and their conjugates, we have
\begin{align}
E_p &= \kap\om_c \int_0^{\infty} d\tau \la a^{\dag}a \ra
+ \frac{g\kap}{2} \int_0^{\infty} d\tau 
\left(\la a^{\dag}\sigma \ra + \la \sigma^{\dag}a \ra\right),
\\
E_e &= g\gamma_p \int_0^{\infty} d\tau 
\left(\la a^{\dag}\sigma \ra + \la \sigma^{\dag}a \ra\right).
\end{align}
Thus, we must evaluate $I_1=\int_0^{\infty}d\tau\la\s^{\dag}\s\ra$,
$I_2=\int_0^{\infty}d\tau\la a^{\dag}a\ra$,
and $I_3=\int_0^{\infty}d\tau\la\s^{\dag}a\ra$.

The equations of motion for $\la\s^{\dag}\s\ra$, $\la a^{\dag}a \ra$, 
and $\la \s^{\dag}a \ra$ are given by
\begin{align}
\frac{d}{dt}\la\s^{\dag}\s\ra &= 
-\gamma \la\s^{\dag}\s\ra -ig(\la \s^{\dag}a \ra-c.c.),
\\
\frac{d}{dt}\la a^{\dag}a \ra &= 
-\kap \la a^{\dag}a \ra +ig(\la \s^{\dag}a \ra-c.c.),
\\
\frac{d}{dt}\la \s^{\dag}a \ra &= 
i(\tom_d^{\ast}-\tom_c) \la \s^{\dag}a \ra +ig(\la a^{\dag}a \ra-\la\s^{\dag}\s\ra).
\end{align}
Integrating these equations with respect to $\tau$,
we obtain $1=\gamma I_1+ig(I_3-I_3^*)$,
$0=\kap I_2-ig(I_3-I_3^*)$,
$0=i(\tom_d^{\ast}-\tom_c)I_3+ig(I_2-I_1)$.
When $\gamma=0$, these equations are solved to yield $I_2=1/\kappa$ 
and $I_3=(i/g)\times(\tom_c^*-\tom_d)/(\tom_c+\tom_d-\tom_c^*-\tom_d^*)$.
Since $E_p=\kap\om_c I_2+(g\kap/2)(I_3+I_3^*)$ 
and $E_e=g\gamma_p(I_3+I_3^*)$, we obtain eqs.~(\ref{EnergyE}) and (\ref{EnergyB}).

%%%%%%%%%%%%%%%%%%%%%%%%%%%%%%%%%%%%%%%%%%%%%%%%%%%%%%%%%%%%%%%%%%%%%%%%%%%%%%%
%%%%%%%%%%%%%%%%%%%%%%%%%%%%%%%%%%%%%%%%%%%%%%%%%%%%%%%%%%%%%%%%%%%%%%%%%%%%%%%


\begin{thebibliography}{99}
\bibitem{footnote}
By using a frequency filtering technique, we can extract one of the two contributions to the purity. 
However, frequency filtering is shown to be ineffective for improving the purity.
\end{thebibliography}
\end{document}